\documentclass[journal,10pt]{IEEEtran}
\usepackage{algorithm,multirow}
\usepackage[noend]{algorithmic}
\usepackage{graphicx}
\usepackage{comment}
\usepackage{cite}
\usepackage{multirow}
\usepackage{amssymb,amsthm,mathrsfs}
\usepackage{amsfonts,epsf,graphicx,times,amsmath,multirow,url,cite}
\usepackage{float}
\usepackage{flexisym}
\usepackage[capitalise]{cleveref}

\usepackage{amsthm}
\usepackage{subcaption}
\usepackage{color}
\graphicspath{{figs/}}
\DeclareGraphicsExtensions{.eps,.pdf,.jpeg,.png}
\usepackage{array}
\newcolumntype{L}[1]{>{\raggedright\let\newline\\\arraybackslash\hspace{0pt}}m{#1}}
\newcolumntype{C}[1]{>{\centering\let\newline\\\arraybackslash\hspace{0pt}}m{#1}\newcolumntype{R}[1]{>{\raggedleft\let\newline\\\arraybackslash\hspace{0pt}}m{#1}}}
\makeatletter
\def\ps@headings{%
\def\@oddhead{\mbox{}\scriptsize\rightmark \hfil \thepage}%
\def\@evenhead{\scriptsize\thepage \hfil \leftmark\mbox{}}%
\def\@oddfoot{}%
\def\@evenfoot{}}
\makeatother
\usepackage [english]{babel}
\usepackage [autostyle, english = american]{csquotes}
\MakeOuterQuote{"}

\IEEEoverridecommandlockouts

\begin{document}

\date{}


\title{\huge Differentiate Quality of Experience Scheduling for Deep Learning Inferences with Docker Containers in the Cloud}
\author{
Ying Mao,~\IEEEmembership{Member,~IEEE,}  Weifeng Yan, Yun Song, Yue Zeng, Ming Chen,~\IEEEmembership{Student Member,~IEEE,} 
\\ Long Cheng,~\IEEEmembership{Senior Member,~IEEE}, Qingzhi Liu,~\IEEEmembership{Member,~IEEE} 
\IEEEcompsocitemizethanks{

\IEEEcompsocthanksitem Y. Mao, W. Yan, Y. Song, Y. Zeng and M. Chen are with the Department of Computer and Information Science at Fordham University in the New York City. E-mail: \{ymao41, wyan16, ysong96, yzeng44, mchen177\}@fordham.edu
\IEEEcompsocthanksitem L. Cheng is with the School of Control and Computer Engineering, North China Electric Power University, Biejing, China. E-mail: lcheng@ncepu.edu.cn
\IEEEcompsocthanksitem Q. Liu is with Information Technology Group, Wageningen University, The Netherlands E-mail: qingzhi.liu@wur.nl

}
}

\newcommand{\sol}{\texttt{DQoES}}

\newtheorem{example}{Example}

\maketitle

\begin{abstract}

With the prevalence of big-data-driven applications, such as face recognition on smartphones and tailored recommendations from Google Ads, we are on the road to a lifestyle with significantly more intelligence than ever before. 
Various neural network powered models are running at the back end of their intelligence to enable quick responses to users. Supporting those models requires lots of cloud-based computational resources, e.g., CPUs and GPUs.
The cloud providers charge their clients by the amount of resources that they occupy. Clients have to balance the budget and quality of experiences (e.g., response time). The budget leans on individual business owners, and the required Quality of Experience (QoE) depends on usage scenarios of different applications. For instance, an autonomous vehicle requires an real-time response, but unlocking your smartphone can tolerate delays. However, cloud providers fail to offer a QoE-based option to their clients. In this paper, we propose \sol, differentiated quality of experience scheduler for deep learning inferences. \sol~ accepts clients' specifications on targeted QoEs, and dynamically adjusts resources to approach their targets.
Through the extensive cloud-based experiments, \sol~ demonstrates that it can schedule multiple concurrent jobs with respect to various QoEs and achieve up to 8x times more satisfied models when compared to the existing system.

\end{abstract}


\section{Introduction}
\label{intro}

In recent years, the dramatic growth of big data has been witnessed from different sources, such as webs, cameras, smartphones, and sensors. To utilize the data, many research and applications have started to be powered by big data analytics in both theory and practice~\cite{zhu2020high, cui2021nonparametric, du2020multiple}. For example, Google AdSense~\cite{adsense} recommends clients the advertisements by studying the data that they generated through Google Workspace. Apple FaceID~\cite{faceid} learns from users' face images with a TrueDepth camera for a secure authentication solution. 
Consequently, various learning algorithms and models are proposed to facilitate the analytical processes. 
In this domain, deep learning technologies, such as artificial neural networks (e.g., convolutional neural network~\cite{cnn} and recurrent neural network~\cite{rnn}), are popular solutions to enhance learning and improve the results. 

Although deep learning algorithms empower many popular applications, obtaining a well-trained model is an intellectual challenge and time and resource-intensive task.
In reality, few companies could afford to collect their datasets and train the models from scratch. Big players, such as Amazon SageMaker~\cite{sagemaker} and Azure Machine Learning~\cite{aml}, provide pre-trained machine learning services with a wide range of productive experiences to build and train learning models. The cloud service of ready-made intelligence makes those techniques practical for limited-budget companies. 

While a well-trained model could be directly obtained by loading a pre-trained one, deploying a model
is still a computationally intensive task. The cloud providers, such as Microsoft Azure~\cite{azure} and Amazon Web Service~\cite{aws}, offer the resources, e.g., CPU and GPU, to host those models. 
In general, clients are charged by the resource levels that they occupy. For example, an a1.xlarge EC2 instance from Amazon Web Service contains 4 vCPUs and 8GB memory and is priced at \$0.1 per hour. A larger instance provides more resources and a faster response to the users. 
Without any constraints, the cloud clients should provide their end-users the fastest-possible service.  
In practice, however, they have to balance the budget from its own end and deliver the quality of experiences (QoEs) on the user's side (e.g., response time). 
The developers have financial budgets to deploy their models. 
More importantly, different applications lead to various QoEs requirements.
For instance, in an autonomous vehicle (e.g., Waymo~\cite{waymo}), the system has to immediately
react to an object that suddenly appears in the front. Milliseconds delay may result in life loss; In an intrusion detection system~\cite{gamage2020deep}, any delayed response may lead to sensitive information leakage;
however, when unlocking the smartphones, users can tolerate
a short delay, up to 1 second~\cite{delay}.


Different applications have various usage scenarios, which naturally result in varied service level objectives and QoE requirements. The developers can survey their targeted end-users to understand the QoE requirements (e.g., response time). However, when deploying their learning models for inferences, they are provided with various resource levels. These levels fail to link to QoEs directly and are hard to quantify. To fill the gap between provided and needed resources, the cloud providers should enable a user-specified QoE and, according to this value,  dynamically adjust resources to satisfy the required QoEs. With this feature, varied applications can be easily differentiated by the targeted QoEs. Due to the shared nature of the cloud and the current business models, it is a challenge to differentiate the deep learning inferences, which require micromanagement on the resources of each task at runtime.

In this project, we propose \sol~\cite{song2020differentiate}, a scheduler that supports user-specified Quality of Experiences. 
When launching a back-end service, a value can be given to \sol~ as a targeted QoE to a specific application. Different applications with various QoEs are hosted by the same cluster and share the resources.  
In a cluster, \sol~ keeps monitoring running models as well as their associated QoE targets and tries to approach their targets
through efficient resource management. The main contributions of this paper are summarized as
follows.

\begin{itemize}

\item We introduce the differentiated quality of experience to the deployment of various deep learning inferences.
With respect to varied budget limitations and usage scenarios, our system is able to react differently.

\item Without modifying the existing cloud system, we propose \sol~ with a suite of algorithms that accepts targeted QoE
specifications from clients
and dynamically adjusts resources to approach their targets to achieve satisfying performance at the user's end.

\item \sol~ is implemented into Docker Swarm~\cite{docker}
, a popular container management toolkit in the industry.
Based on extensive 
cloud-executed 
experiments, \sol~ demonstrates
its capability to react to different workloads and achieves up to 8x times 
more satisfied models when compared to the existing system.

\end{itemize}

The rest of this paper is organized as follows. In Section \ref{rel},
we report the related work. In Section \ref{sys}, we introduce our
\sol~ system design. We present the proposed
\sol~ algorithms in Section \ref{sol}.
We carry out extensive cloud-based evaluation of \sol~ in Section \ref{eval}
and conclude this paper in Section \ref{con}.


\section{Related Work}
\label{rel}

Learning from massive datasets to drive businesses is drastically reshaping our economy.
Lots of research have been done to improve the performance of the learning models
from various perspectives, such as accuracy~\cite{xie2019self, xie2020self, sablayrollesgoing} and model size~\cite{ouahabi2021deep, chen2019eyeriss, sun2017revisiting}.

Although promoting the accuracy of a learning model is important, obtaining a well-trained model is still both a time and resource-consuming process. Typically, deep learning models are trained and hosted on the cloud. Focusing on the cloud platform, many research projects have been done to improve resource utilization. Varys~\cite{chowdhury2014efficient} aims to reduce the communication cost in data-parallel applications, such as Hive and MapReduce. Given the flow information (e.g., sizes and endpoints), it utilizes a two-step algorithm to minimize coflow completion times and allocates minimum required resources. PIAS~\cite{bai2015information}, however, does not require any prior knowledge about flows in the data center network. It leverages multiple priority queues to implement a multiple-level feedback queue. With the feedback queue, PIAS flow is gradually demoted from higher priority queues to lower-priority queues based on the number of bytes it has sent. Varys and PIAS improve targeted applications significantly. However, they fail to be directly adopted to deep learning applications, which are computationally intensive rather than network intensive.   

Focusing on deep learning applications, many ongoing research projects have been proposed to accelerate the training phase in a cloud environment~\cite{mao2017draps, zheng2019flowcon, mao2021speculative, fu2019progress, huang2020lightweight, luo2020towards, peng2018optimus}. For example,  FlowCon~\cite{zheng2019flowcon} enables dynamic resource management with respect to the growth efficiency of the loss function. From the system point of view, ProCon~\cite{fu2019progress} optimizes the cluster by selecting a best-fit host for incoming tasks. 
In order to better utilize the resources, Optimus~\cite{peng2018optimus} employs an online fitting technique to predict model convergence and sets up performance models to estimate training speed. 
However, these studies focus on the deep learning training phase, and proposed techniques cannot be adopted directly to the inferences. In addition, they fail to take a view from users' experiences.

From the perspective of deep learning inferences, several works take the quality of experience from the end-users into account~\cite{gujarati2020serving, qiu2020firm, berral2020ai4dl, zheng2019target, romero2021infaas, cui2021enable, ogden2021pieslicer, xu2021talos, wang2021morphling}. For example, Clockwork~\cite{gujarati2020serving} exploits predictable execution times of deep neural network inference. It attempts to reduce the latency by ordering users' queries based on their Service Level Objectives (SLOs) and only executing one request at a time. While Clockwork improves the system throughput and predictability, it does not consider the cloud as a multitenancy system. FIRM~\cite{qiu2020firm} considers the resource sharing. It reduces the SLO violations by leveraging online telemetry data and utilizing a support vector machine based model to identify the key components that are likely to become resource bottlenecks. While FIRM improves performance predictability, it assumes that cloud machines are equipped with unlimited resources, which is unrealistic.
INFaaS~\cite{romero2021infaas} proposes an automated model-less
system for distributed inference serving. Instead of selecting a specific model, developers are able to specify the performance and accuracy requirements. However, INFaaS focuses on the model auto-selection and fails to guide the container configuration. 
To better understand the resource usage pattern, AI4DL~\cite{berral2020ai4dl, buchaca2020proactive} studies the training processing of deep learning applications within IBM's DLaaS containers. It demonstrates that CPU and memory usage can be reduced significantly by leveraging typical resource usage patterns. Despite improvement observed, AI4DL requires statistical data from previous trials.   
Bringing users into the loop, TRADL~\cite{zheng2019target} tries to differentiate the applications by taking advantage of a user-specified target. However, it only focuses on the training process and the specified objectives cannot be adopted to the deep learning inference.

While efforts have been devoted from different perspectives, considering various usage scenarios, limited works focus on differentiating QoEs for deep learning backed inferences. 
Considering the end-users, we propose \sol, which supports a user-specified QoE targets, such as reaction time. Given a multitenancy cloud system, \sol~ is able to approach specific targets through dynamic resource configuration.

\section{\sol~ System Design}
\label{sys}

This section presents the system architecture of \sol~ in detail, including its framework, design logic and functionalities of key modules.

\subsection{\sol~ framework}
A typical management system for a cluster of containers, such as Docker Swarm~\cite{docker} and CNCF Kubernetes~\cite{k8s}, involves managers and workers. It works in a distributed manner with managers and workers. The managers interact with the users, analyze their requests, and manage the workers associated with this cluster and workers. The worker nodes are the workhorses that provide the computing resources, such as CPU, GPU, and memory, to execute the tasks, store the data, and report their status to managers.

\sol~ employs the manager-worker architecture. In \sol, the manager collects QoE targets from clients, passes them to the workers, and maintains an overall performance assessment table to monitor the real-time performance of all the active workloads in the system.
The workers, who host the deep learning applications, track the performance of each individual model and its real-time resource usages. According to the QoE targets, the worker dynamically adjusts resource limits for each container to achieve the best overall performance among all models that resident in it.

\begin{figure}[ht]
\centering
\includegraphics[width=0.95\linewidth]{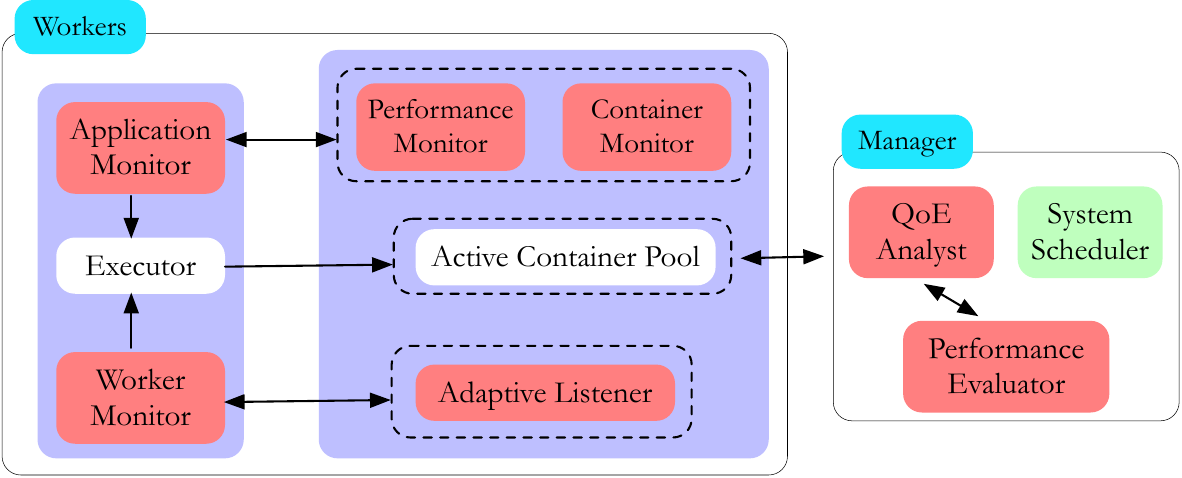}
\caption{\sol~System Architecture}
\label{fig:system}
\end{figure}

\subsection{\sol~ Modules}

As demonstrated in Figure~\ref{fig:system}, \sol~ consists of four major modules, an Application Monitor, a Worker Monitor, and an Executor on the worker side. A QoE Analyst on the manager side. 
Each module runs independently and exchanges information about models inside the containers as well as the worker status. Their functionality is detailed below.

{\bf Application Monitor}: it maintains performance metrics for each deep learning model that runs this worker.
For example, when users query the model, the response time would be recorded along with the resource usage associated with the time cost. Compared with the QoE targets, 
The \sol~ executes an algorithm to adjust resource assignments for active containers and approaches the targeted values iteratively.

{\bf Worker Monitor}: it measures the active container pool. When a manager assigns a new model, the worker monitor adds it to the pool and keep tracking performance difference in each iteration by using an adaptive listener. 
The listener hosts an algorithm that regulates the frequency of updating the resource assignment.

{\bf Executor}: it is a key module that accepts the workload and collects data on the
worker. Based on data, it calculates the parameters required by the algorithms (described in Section~\ref{sol}) in Application and Worker Monitor. Upon receiving a new plan for resource configuration for each container, the Executor will interrupt the current limits and update containers with newly calculated values.

{\bf QoE Analyst}: it resides on the manager's side. A QoE analyst interacts with clients and collects the QoE targets for each of the deep learning models. The targeted values are sent to workers through the System Scheduler.
When the status reports arrive from workers, it utilizes a Performance Evaluator to preserve data and monitor the overall
system performance.

\subsection{System Optimization Problem}

In \sol, we consider deep learning applications host on the cloud with a cluster of containers, and each learning model resides in a container. \sol~ aims to provide users the service with a predefined QoE target.

In the system, we denote $c_i \in C$ to be a container that serves one deep learning application.  
For each $c_i$, we have $r_i \in R$ is the resource usage of $c_i$ that can be monitored through the 
system APIs, such as \texttt{docker stats}, where function $R(c_i, t) = r_i$ is defined to retrieve the resource usages
in real-time. 

For the model that runs in each container, we keep tracking its performance $P (c_i, t) = p_i$, which
is a value to represent a predefined quality metric, e.g. response time, of a particular service. 
Additionally, there is an pre-defined QoE targeted value for each model, $o_i \in O$.
Then, the quality of a given container $c_i$ (a learning model runs inside) is $q_i = o_i - p_i$.
Therefore, the system performance on a specific worker node $W_i$ is, 
\begin{equation}
Q_{W_i} = \sum_{c_i \in W_i} q_i 
\end{equation}

\sol~ anticipates a system that minimizes the performance difference between the predefined QoE target and current
container outputs. 
Considering the real-time container outputs, \sol~ classifies running models into three different categories. 

\begin{itemize}
\item Class G: The models in this category perform better than the preset QoE value, e.g. response faster than the target.

\item Class S: The models in this category achieve the predefined QoE targets and they are satisfied containers.

\item Class B: The model in this category is underperformed due to lacking of resources or unrealistic targets. 
\end{itemize}

Assume that there are n containers, each runs one learning model in the system. Then, our
performance optimization problem, $\mathcal{P}$, can be formalized as

\begin{align}
\label{constraint_eqn}
&&&\mathcal{P}:\text{Min}\sum_{c_i \in B ~OR~ c_i \in G}^{n} q_i \\  \nonumber
&&&\textbf{s.t.} \sum_{c_i \in W_i}^{n} r_i \leqslant T_{R} \quad  \\ \nonumber
\end{align}

The following table summarizes the parameters that utilize in \sol. Note that due to the variety of user-specified objectives, currently, \sol~ does not consider the fairness among the running containers. When dynamically adjusting the resources, the objective is to maximize the number of containers that satisfy the targeted \sol.

\begin{table}[ht]
	\centering
	\caption{Notation Table}
	\scalebox{0.9}{
		
		\begin{tabular}{ | c | c |  }				
			\hline 
			$W_i$ & Worker $i$ \\ \hline
			$c_{i} \in C$      & The container $i$ in the set $C$  \\ \hline	
			$G$ &  Set $G$ contains the $c_i$ with a better performance than $o_i$ \\ \hline
			$S$ &  Set $S$ contains the $c_i$ with targeted performance $o_i$ \\ \hline
			$B$ &  Set $B$ contains the $c_i$ with a worse performance than $o_i$ \\ \hline
			$o_i \in O$ & Objective value $o_i$ for $c_i$ in the set $O$ \\ \hline
			$r_i \in R$ & Resource usage $r_i$ for $c_i$ in the set $R$  \\ \hline
			$R(c_i, t)$ & The function that uses to calculate $r_i$ at time $t$ \\ \hline
			$p_i$ & The performance indicator of $c_i$ \\ \hline
			$P (c_i, t)$ & The performance function that calculate $p_i$ at time $t$ \\ \hline
			$q_i$ & The quality of $c_i$ that can be calculated by using $o_i$ and $p_i$ \\ \hline
			$T_R$ & The total resource usage of containers on a worker $i$ \\ \hline
			$R_G, R_S, R_B$ & The resource usage of containers in set $G$, $S$ and $B$ \\ \hline
			$Q_G, Q_S, Q_B$ & The sum of qualities for containers in set $G$, $S$ and $B$  \\ \hline
			$L(c_i, t)$ & The limit of the resource usage for container $c_i$ at time $t$ \\ \hline		
		\end{tabular}	
	}	
	\label{table:notation}
\end{table}




	

%


\section{\sol~ Solution Design}
\label{sol}

This section presents the detailed design of the performance management algorithms in \sol, which can adjust resource assignment for containers at run time.
In addition, we discuss the algorithm of an adaptive listener to reduce the overhead of \sol.

\begin{algorithm}[!t]
\caption{Performance Management on $W_i$}
\begin{algorithmic}[1]
\STATE  Initialization: $W_i$, $c_i \in C$, $o_i \in O$, $p_i \in P$,

\item[]

\FOR {$c_i \in W_i$}
\STATE $R(c_i, t) = r_i$

\STATE $P(c_i, t) = p_i$
\STATE $q_i = o_i - p_i$

\IF{$q_i > \alpha \times o_i$}
\STATE $G\text{.insert}(c_i)$
\STATE $Q_G = q_i + Q_G$
\STATE $R_G = r_i + R_G$
\ELSIF {$q_i < - \alpha \times o_i$}
\STATE $B\text{.insert}(c_i)$
\STATE $Q_B = q_i + Q_B$	
\STATE $R_B = r_i + R_B$	
\ELSE
\STATE $S\text{.insert}(c_i)$

\ENDIF		
\ENDFOR



\item[]
\FOR {$c_i \in W_i$}
\IF {$c_i \in G$}
\STATE $L(c_i, t+1) = L(c_i, t) * (1 - \frac{q_i}{Q_G}\times R_G \times \beta)$
\IF {$L(c_i, t+1) < \frac{1}{2\times |C|}$}
\STATE $L(c_i, t+1) =\frac{1}{2\times |C|}$
\ENDIF
\ELSIF {$c_i \in B$}
\STATE $L(c_i, t+1) = L(c_i, t) * (1 + \frac{q_i}{Q_B} \times R_G \times \beta)$
\IF {$L(c_i, t+1) > T_R$}
\STATE $L(c_i, t+1) = T_R$
\ENDIF
\ENDIF
\ENDFOR		

\item[]
\end{algorithmic}

\label{alg:1}
\end{algorithm}

\subsection{\sol~ Performance Management}
\sol~manages the QoE of learning models through dynamically adjusting the resource distribution of their host containers.
The administrators of a container management system, such as Docker Swarm, are able to configure a "soft limit" for a running container. The limit specifies an upper bound of the resources that a container can be assigned. For example,
the command \texttt{docker update --cpus="1.5" Container-ID} can set the container is guaranteed at most one and a half of the CPUs, assuming the host has at least 2 CPUs.

\sol~ utilizes Algorithm~\ref{alg:1} to manage the performance, in terms of QoE targets, of each deep learning application.
Firstly, it initializes the parameters required by the algorithm (Line 1).
For every active container, it fetches the runtime resource usage, $r_i$, the current performance value, $p_i$ and, together with the predefined target, it calculates the quality of the container, $q_i$, at this moment (Line 2-5).

\sol~, then, classifies all learning models into three different categories, $G$, $B$, and $S$.
Based on the current quality of $c_i$, if it is larger than $\alpha \times o_i$, which means the quality level is higher than the target, $c_i$ is marked as $G$. The $\alpha$ is a percentage value that is used to represent the developer's tolerance of the target. If $q_i$ is smaller than $ - \alpha \times o_i$, it indicates the quality level is lower than the target, $c_i$ is set to be $B$. In the case that $q_i$ falls within the interval $[-\alpha\times o_i, \alpha \times o_i]$, it suggests that the service provided by $c_i$ satisfies the predefined QoE target and $c_i$ joins $S$. When classifying the containers, \sol~ calculates a total quality value of learning models in $G$ and $B$ as well as the sum of resource usages, respectively (Line 6-15).

For the containers in $G$, which perform better than their QoE targets, Algorithm~\ref{alg:1} cuts their resource usages to reduce the performance and approach their targets. For underperformed containers in $B$, \sol~ tries to increase their resource allocation to help them move toward the QoE targets. Precisely, the following two branches are executed.

\begin{itemize}

\item For $c_i \in G$, the total resource limits are reduced from the previous values by $R_G \times \beta$, where $\beta$ is a parameter that the system administrator can configure to update the containers gradually.
The degree of the reduction depends on how far away a $c_i$ is from its QoE target. For example, when a learning model is
performing slightly better than the targeted value, the value of $\frac{q_i}{Q_G}$ is very small, and the resource reduction is limited (Line 16-18). When updating the limits, \sol~ set $\frac{1}{2\times |C|}$ to be the lower bound of the model to prevent abnormal behaviors due to lacking resources (Line 19-20).

\item For containers in $B$, the saved resources from $G$ will be reallocated to them. Similar to $G$, the degree of increase depends on how bad a container performs when compared to its QoE target. For example, if a given container, $c_i$, is underperforming a lot, which results in a large value of $\frac{q_i}{Q_B}$ and leads to a hike in the resource limit (Line 21-22). However, there is an upper limit (e.g., hardware limits) of the resource allocation, $T_R$, to each container (Line 23-24).   
 
\end{itemize}

\subsection{Adaptive Listener on \sol}

Algorithm~\ref{alg:1} aims to dynamically update the resource allocation for active containers in order to make the learning models approach their predefined QoE targets iteratively. To achieve the goal, \sol~ has to keep measuring performance of each container and their resource usages, which create overhead to the system. In addition, with a fixed number of active containers and untouched QoE targets, \sol, after several iterations, converges to the state with a stable resource distribution. 
Therefore, it is unnecessary to frequently collect information and execute Algorithm~\ref{alg:1}.

In \sol, it implements an adaptive listener with an exponential back-off scheme to control the frequency.
Algorithm~\ref{alg:2} presents the details in the listener. 
At the very beginning, it initializes the parameters, such as the sets $G, S, B$, that require for execution (Line 1).
Then, the sum of qualities, $Q_G, Q_B, Q_S$ for each of the container set is calculated. Please note that, for $c_i \in S$, the learning models have satisfied their QoE targets. Thus, their $q_i = 0$ and we use the number of containers in $S$ to represent
$Q_S$. While $Q_G, Q_B, Q_S$ are per iteration values, \sol~ uses $Q_G(t), Q_B(t), Q_S(t)$ to record the time serial of these values (Line 2-11). 

Then, \sol~ compares the most recent two values of $Q_G$ and $Q_B$. If both $Q_G$ and $Q_B$ are approaching to 0, which indicates by $Q_G(t+1) < Q_G (t) $ and $Q_B (t+1) > Q_B(t)$, it means they are all converging to $S$. When the trend maintains for 3 consecutive iterations, Algorithm~\ref{alg:2} doubles the interval, which reduce the frequency of updating the resource allocation (Line 12-16).  However, when $Q_S(t)$ reduces, it suggests the stable state is broken. It could be abnormal usages of one particular container or a new one joins the system.  In this case, \sol~ resets the interval and execute Algorithm~\ref{alg:1} immediately (Line 17-20). When the system performance is still bouncing, \sol~ maintains the original interval (Line 21-23).

\begin{algorithm}[!t]
\caption{Adaptive Listener on $W_i$}
\begin{algorithmic}[1]

\STATE  Initialization: $W_i$, $c_i \in W_i$, $o_i \in O$, $p_i \in P, q_i, G, S, B$, $IV = IV_{initial}$

\item []

\FOR {$c_i \in W_i$}
	\IF {$c_i \in G$}
		\STATE $Q_G = Q_G + q_i$
		\STATE $Q_G (t) = Q_G$
	\ELSIF {$c_i \in B$}
		\STATE $Q_B = Q_B + q_i$	
		\STATE $Q_B (t) = Q_B$
		
	\ELSIF {$c_i \in S$}
		\STATE $Q_S = |S|$
		\STATE $Q_S (t) = Q_S$
	\ENDIF
\ENDFOR	

\item []

\IF {$ |Q_G (t+1) < Q_G (t)|$ \& $|Q_B (t+1) > Q_B (t)|$}
	\STATE $i++$
	\IF {$i \geqslant 2$}
		\STATE $IV = IV \times 2$
		\STATE $i = 0$
	\ENDIF
\ELSIF {$Q_S (t+1) < Q_S(t) $}
	\STATE 	$IV = IV_{initial}$
	\STATE $i=0$
	\STATE Execute Algorithm 1
\ELSE
	\STATE $IV = IV$	
	\STATE $i=0$
\ENDIF

\end{algorithmic}
\label{alg:2}
\end{algorithm}
\section{Performance Evaluation}
\label{eval}

In this section, we evaluate the effectiveness and efficiency of \sol~ through intensive, cloud-executed experiments. 

\subsection{Experimental Framework and Evaluation Metrics}

\sol~ utilizes Docker Engine~\cite{engine} 19.03 and is implemented as a plugin module that runs on both local and cluster versions. We build a testbed on NSF Cloudlab~\cite{cloudlab}, which is hosted by the Downtown Data Center - University of Utah. Specifically, the testbed uses the M510 physical node containing Intel Xeon D-1548 and 64GB ECC Memory. 

\sol~ is evaluated with various deep learning models
using both the Pytorch and Tensorflow platforms. 
When conducting experiments, pretrained parameters from the platforms are loaded
with the built-in workloads. 
The time for each image recognition task is far less than 1 second. However, the cost for real-time 
reconfiguration in \sol~ fails to take action in seconds level since the more frequent it updates the system,
the more overhead it introduces. Therefore, \sol~ utilizes batch processing and defines 100 images as a batch.
Table~\ref{table:workload} lists the models used in the experiments.

\begin{table}[ht]
	\centering
	\caption{Tested Deep Learning Models}
	\scalebox{0.9}{
		
		\begin{tabular}{ | c | c | }				
			\hline
			Model~\cite{pytorchexample}\cite{tensorflowexample}  & Platform  \\ \hline	
			Visual Geometry Group (VGG-16) 	 & P/T   \\ \hline	
			Neural Architecture Search Network (NASNetMobile)       &  P/T  \\ \hline
			Inception Network V3  & T \\ \hline
			Residual Neural Network (Resnet-50)  & P \\ \hline
			Extreme version of Inception (Xception)   & T \\ \hline
		\end{tabular}	
	}	
	\label{table:workload}
\end{table}

There are two system parameters in \sol, (1) $\alpha$, the threshold for classifying each job into different sets, G (outperform), B (underperform) and S (satisfied); (2) $\beta$, the value that controls the amplitude of resource adjustments at each round. Due to the page limit, we omit the discussion of these parameters. In the evaluation, we set $\alpha$ = $\beta = 10\%$. Furthermore, the initial value of IV is set to 10 seconds. Note that the system administrator can easily update these values. 

We consider the key metric in \sol~ the number of jobs in S, G, and B. If all the containers are in S, the system has achieved the user-specified objectives on each individual job and delivered the best quality of experience according to the clients. Thus, the difference between objectives and experiences is 0, $Q_S = 0$.
It is possible that the user inputs an unrealistic objective, e.g. exceed theoretical boundaries. In this scenario,
\sol~ attempts to achieve the best possible experience to approach the objective. Therefore, metrics that we used to assess
\sol~ is the gap between predefined objective and real delivered experience that is presented by $Q_S, Q_G, Q_B$, the sum qualities for containers in set $G$, $S$ and $B$.

Before any experiments, we first run each model individually and let it occupy all the resources. With this non-competitive environment, we record the fast-possible response time for each model. The values below this value are unachievable, and others are achievable targets. When selecting the targets for the experiments, unless otherwise specified, we randomly generate both types of the targets to fully understand the performance of \sol.

We evaluate \sol~ on individual severs and a cluster with four nodes. In addition, \sol~ is tested with the following job submission schedules.

\begin{itemize}
\item Burst schedule: it simulates a server with simultaneous workloads, which is challenging for \sol~ to adjust resources according to each objective.

\item Fixed schedule: the time to launch a job is controlled by the administrator. When a new job joins the system, \sol~ will have to redistribute the resources to accommodate it.

\item Random schedule: the launch times are randomized to simulate random submissions of jobs by users in a real cluster. \sol~ has to adjust resources to achieve the best overall experience frequently.
\end{itemize} 

The default Docker platform does not support user-specified targets. In  the  experiments with  cluster settings, we mainly focus on the number of satisfied containers (set $S$) in the whole cluster. The more containers that are categorized to set $S$ indicates the system satisfied more of its clients.  

\subsection{Single Model}
We conducted a set of experiments with a single model.
Each job provides services inside a container with a specific objective.
Our single model experiments use Resnet-50 as the source image. 

\subsubsection{Identical Objectives} 
\begin{figure}[ht]
\centering
      \includegraphics[width=0.90\linewidth]{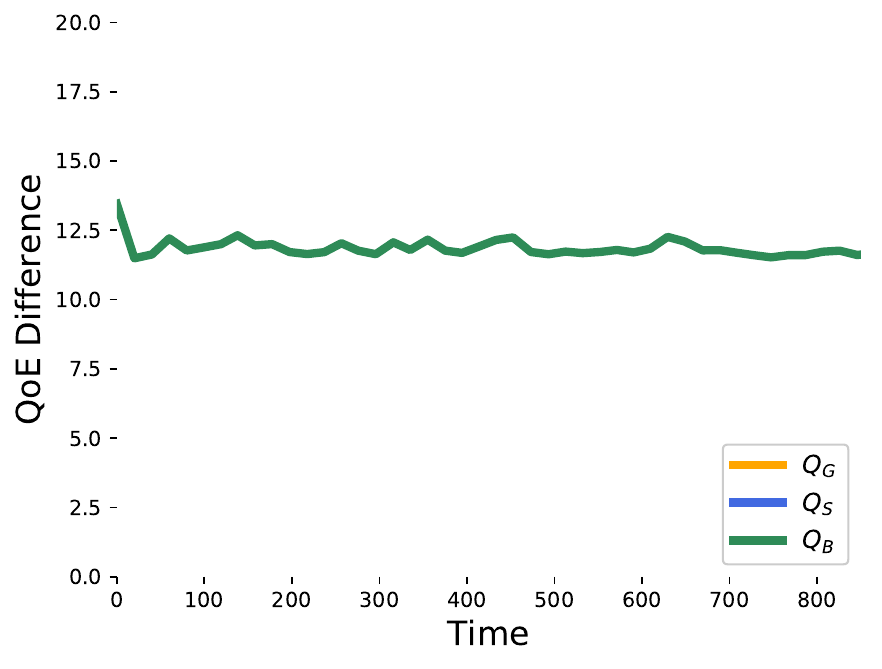}
\caption{Delivered QoE: 10 jobs with the same unachievable objectives (Burst schedule)}
      \label{fig:single:20}
\end{figure}

\begin{figure}[ht]
\centering
      \includegraphics[width=0.90\linewidth]{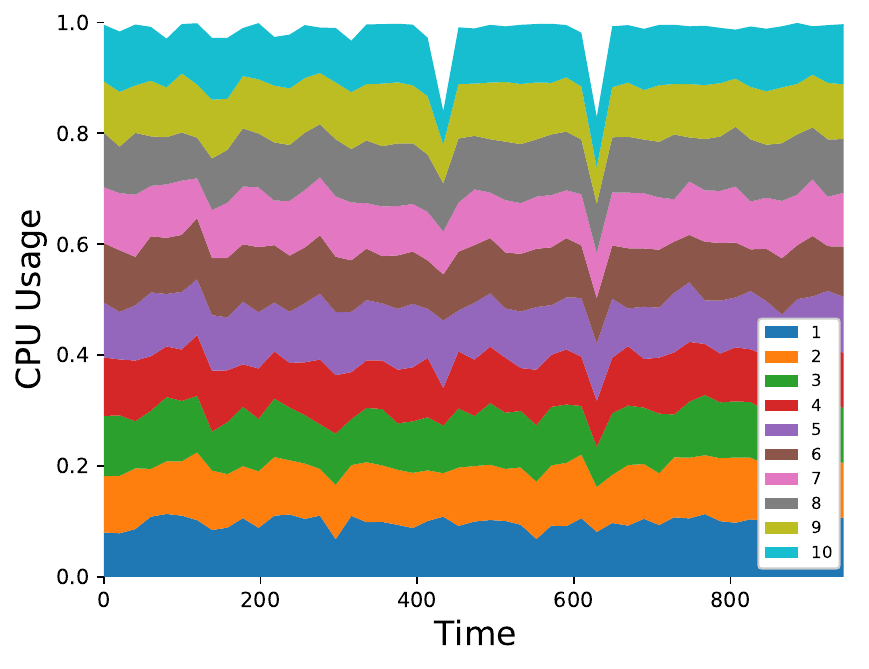}
\caption{CPU distribution: 10 jobs with the same unachievable objectives (Burst schedule)}
      \label{fig:single:20:cpu}
\end{figure}

Firstly, we launch 10 models simultaneously to simulate the burst traffic.
The objectives of them are set to the same value, 20s.  It means that making decisions for a batch of tasks takes 20 seconds.
Fig.~\ref{fig:single:20} presents the QoE from the user's side.
As we can see from the figure, all containers are classified in set B, which means they are all underperformed. This is due to the fact that, given the resources, the objective of 20 seconds
per batch is unachievable. The system evenly distributes all the available resources, but the average value of their QoEs
is still underperformed. For example, at time 350s, the average delivered QoEs is 31.61s with a 0.35 standard deviation.
Fig.~\ref{fig:single:20:cpu} illustrates the CPU distribution among 10 models. 
Given an identical, unachievable objective, the best that \sol~ could do is to approach the targets by evenly distributing all the resources.

\begin{figure}[ht]
\centering
      \includegraphics[width=0.90\linewidth]{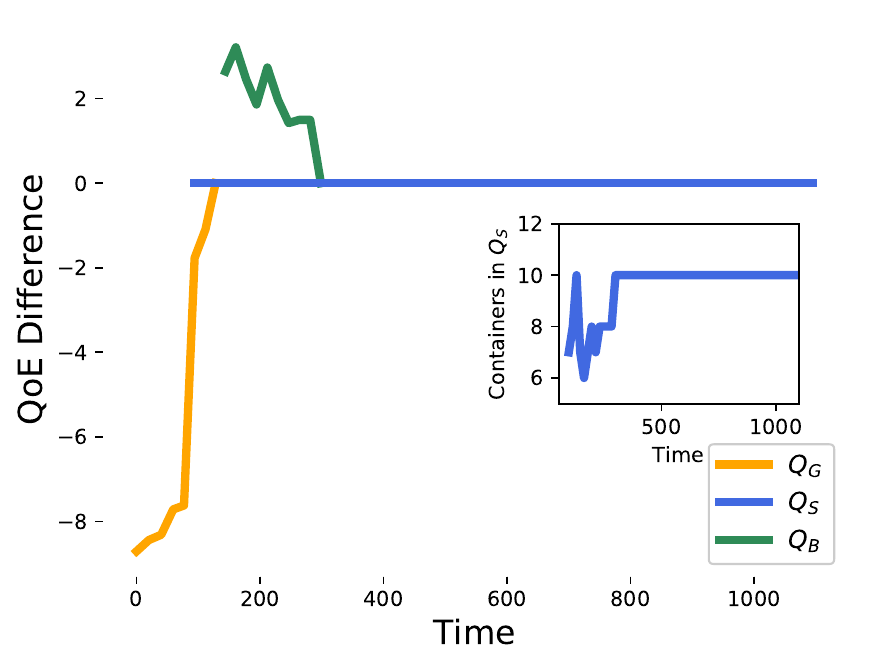}
\caption{CPU distribution: 10 jobs with the same achievable objectives (Burst schedule)}
      \label{fig:single:40}
\end{figure}

\begin{figure}[ht]
\centering
      \includegraphics[width=0.90\linewidth]{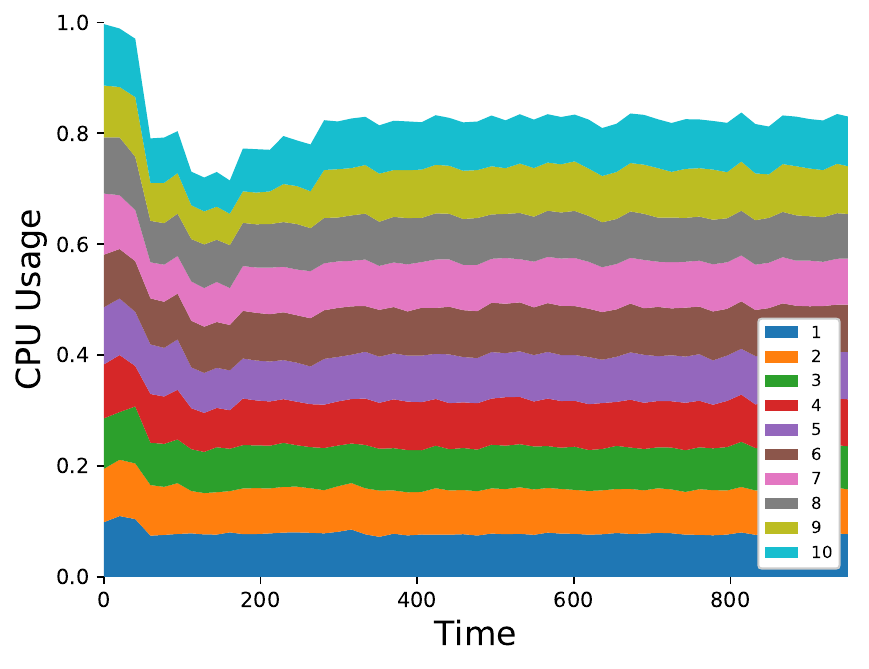}
\caption{Delivered QoE: 10 jobs with the same achievable objectives (Burst schedule)}
      \label{fig:single:40:cpu}
\end{figure}

Next, we utilize the same setting but update the identical objective values to 40, which makes it achievable by the system.
Fig.~\ref{fig:single:40:cpu} presents a more diverse result.
With a dynamic resource configuration, running containers are classified into different sets. For example, at the very beginning,
all of the 10 models are in set G, which means that they all perform better than the predefined objective and thus, occupy more resources than necessary. Consequently, \sol~ starts reducing their resource limits to approach the target, 40s.
At time 94.06s, running models in Container-1, 3, 6, 7, 8, 9, and 10, produce a deliverable experience, which falls within 
$1 \pm\alpha \times$ objective ($\alpha=10\%$ and $o_i = 40$). Therefore, they are classified to set S, which indicates they satisfied the objective. Whenever there is at least one container in S, the $Q_S$ with QoE $=$ 0 shows on the figure.
\sol~ continues to update the resource allocation to improve QoEs of the other three models. However, at time 144.05s, 
container-7, 9, and 10 become underperformed with a batch cost at 47.57s, 50.91s, 47.93s.
The reason is that \sol~ adjust the resource adaptively, and in the previous round of adjustment, it cuts too many resources
from them. As the system goes, \sol~ algorithms converge, and all models achieve their targeted objectives.
The small picture in Fig.~\ref{fig:single:40} illustrates the number of containers in set S. We can clearly see the value goes up and down due to the adaptive adjustment and stabled at 10 for the best performance.
Fig.~\ref{fig:single:40:cpu} presents the real-time resource distribution among all the models. In general, they obtain the same shares between each other due to the identical objective settings. The difference between Fig.~\ref{fig:single:20:cpu} and
Fig.~\ref{fig:single:40:cpu} is that when the system stabilized, the experiments with achievable objectives have resources to accommodate for more workloads. 

\subsubsection{Varied Objectives}

Next, we conducted a set of experiments with varied objectives. The values of objectives are randomly selected 
and consists of both achievable and unachievable targets, which generates more challenges for \sol.

\begin{figure}[ht]
\centering
      \includegraphics[width=0.90\linewidth]{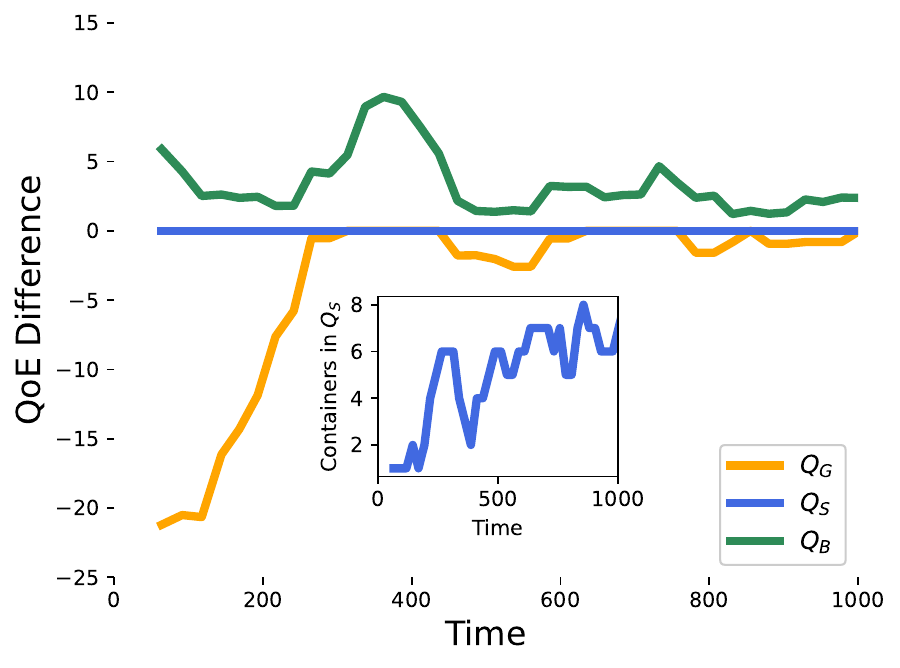}
\caption{CPU distribution: 10 jobs with varied objectives (Burst schedule)}
      \label{fig:single:dif:same}
\end{figure}

\begin{figure}[ht]
\centering
      \includegraphics[width=0.90\linewidth]{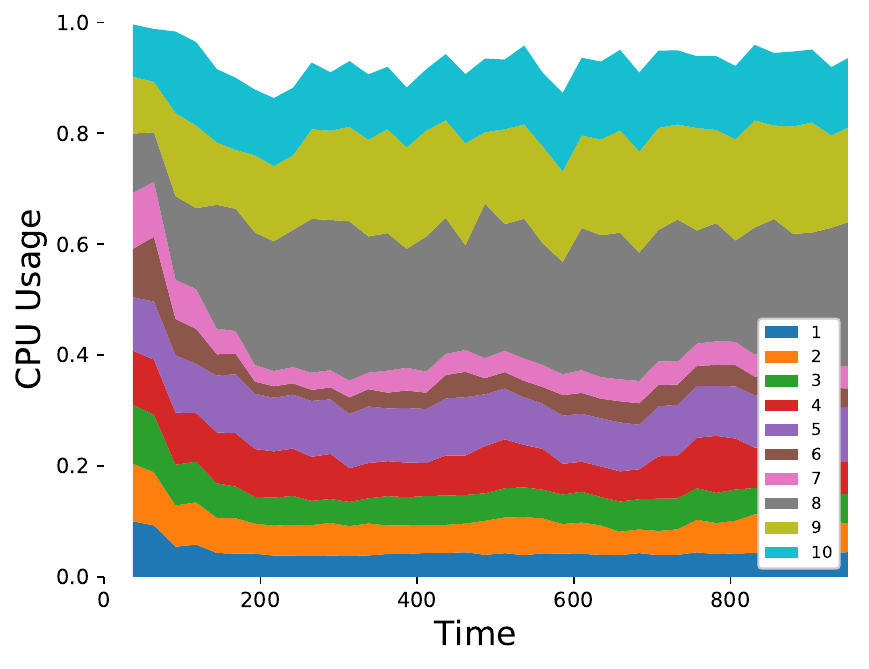}
\caption{Delivered QoE: 10 jobs with varied objectives (Burst schedule)}
      \label{fig:single:dif:same:cpu}
\end{figure}

Fig.~\ref{fig:single:dif:same} and Fig.~\ref{fig:single:dif:same:cpu} plot the results from a burst schedule experiment.
It clearly indicates that \sol~ is able to redistribute resources to approach their objectives individually. At meanwhile, 
\sol~ achieves best overall QoE.  
For example, the objective values in this experiments are 75, 53, 61, 44, 31, 95, 82, 5, 13, 25 for container 1-10, respectively.
Obviously, target value 5 from container-8 is unachievable. \sol~ attempts to approach this objective by allocating more resource
to it than the others (Fig.~\ref{fig:single:dif:same:cpu}).
In Fig.~\ref{fig:single:dif:same}, the number of containers in set $S$ is stabilized at 7 instead of 8 is because \sol~ tries to minimize overall system-wide QoE.

\begin{figure}[ht]
\centering
      \includegraphics[width=0.90\linewidth]{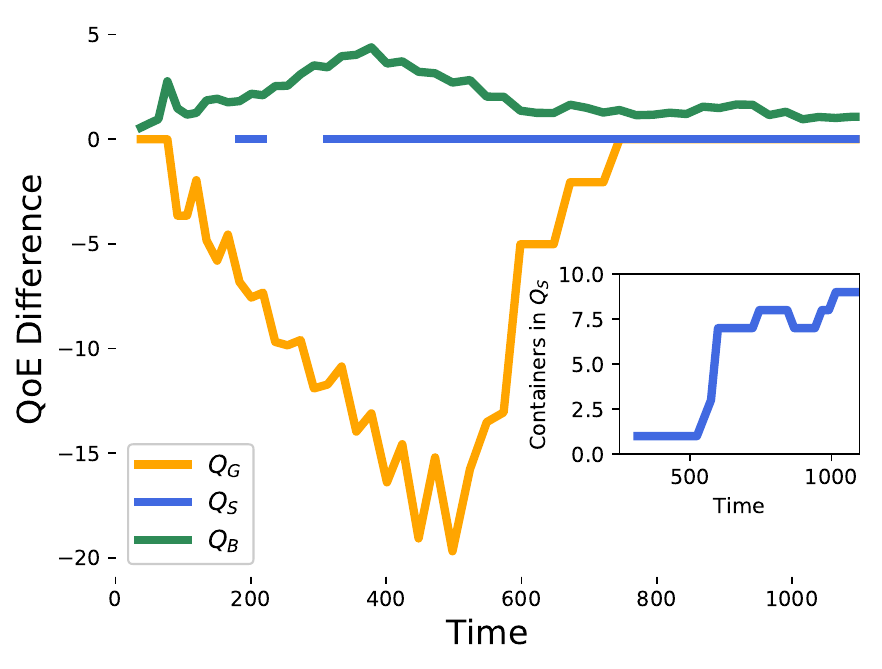}
\caption{Delivered QoE: 10 jobs with varied objectives (Fix schedule)}
      \label{fig:single:dif:gap50}
\end{figure}

\begin{figure}[ht]
\centering
      \includegraphics[width=0.90\linewidth]{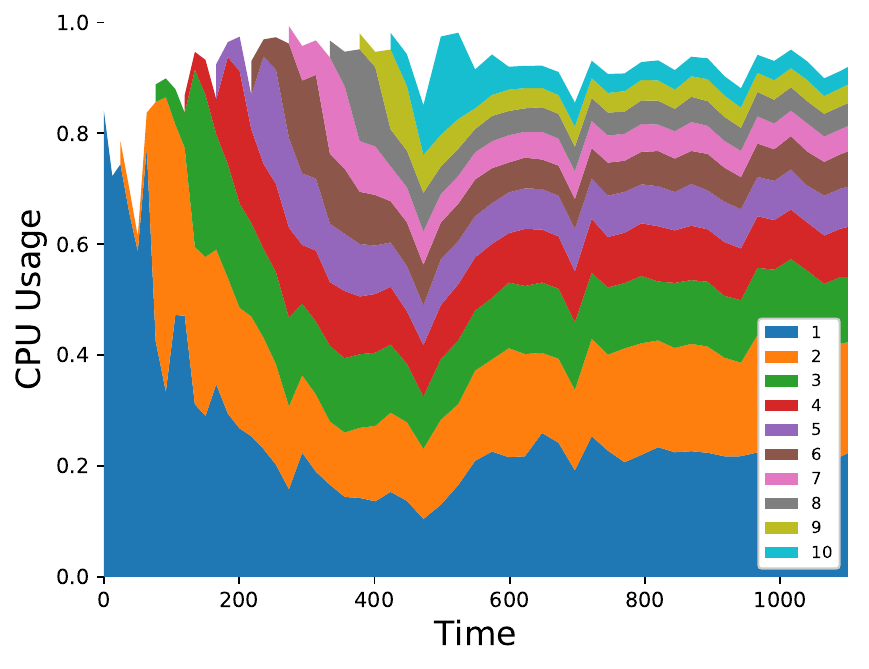}
\caption{CPU distribution: 10 jobs with varied objectives (Fix schedule)}
      \label{fig:single:dif:gap50:cpu}
\end{figure}

Then, we randomly generate a set of objectives and start their host containers with a fix interval of 50 seconds.
Fig.~\ref{fig:single:dif:gap50} shows the result of this experiment.
Different from the previous experiments, the values of $Q_B$ and $Q_G$ change rapidly.
Due to the submission gap of 50 seconds, \sol~ has to update the resource distribution continually.
Whenever a new container joins the system, \sol~ adjusts the allocation to adapt to the new workload and its objective.  
Since the submission happens from 0 to 450 seconds, \sol~ begins convergence after 450 seconds.
At time 769.69s, the number of containers in set $G$ becomes 0, which suggests that all the outperformed models have released 
resources to promote underperformed containers. 
As the system runs, the number of containers in $S$ raises to 8. Only container-1 and 2 with unachievable targets are still in the underperformed set $B$. 
Fig.~\ref{fig:single:dif:gap50:cpu} verified the results from the resource allocation aspect that container-1 and 2 receive a larger amount of resources than others.

\subsection{Multiple Models}

In this subsection, we conduct experiments with multiple models. 
It creates even more challenges for \sol~ since different models have various resource usage patterns and diverse
achievable targets.
When running the experiments, we first randomly select a model image from Table~\ref{table:workload} and assign
an arbitrary number as its objective value. Then, each container randomly picks up a submission time from a given interval.

\subsubsection{Single Node Environment}

\begin{figure}[ht]
\centering
      \includegraphics[width=0.90\linewidth]{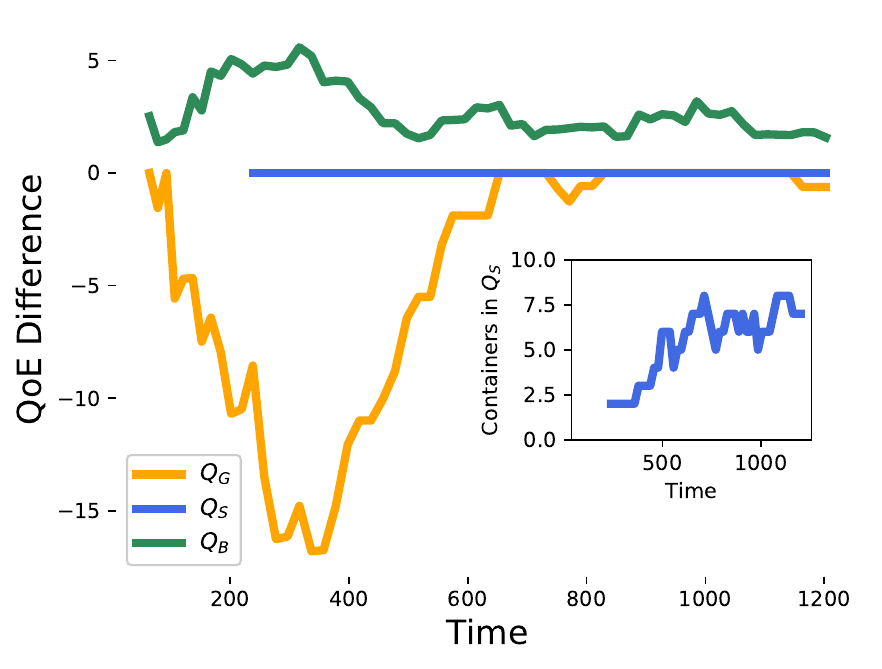}
\caption{Delivered QoE: 10 jobs with varied objectives (Random schedule)}
      \label{fig:mul:random}
\end{figure}

\begin{figure}[ht]
\centering
      \includegraphics[width=0.90\linewidth]{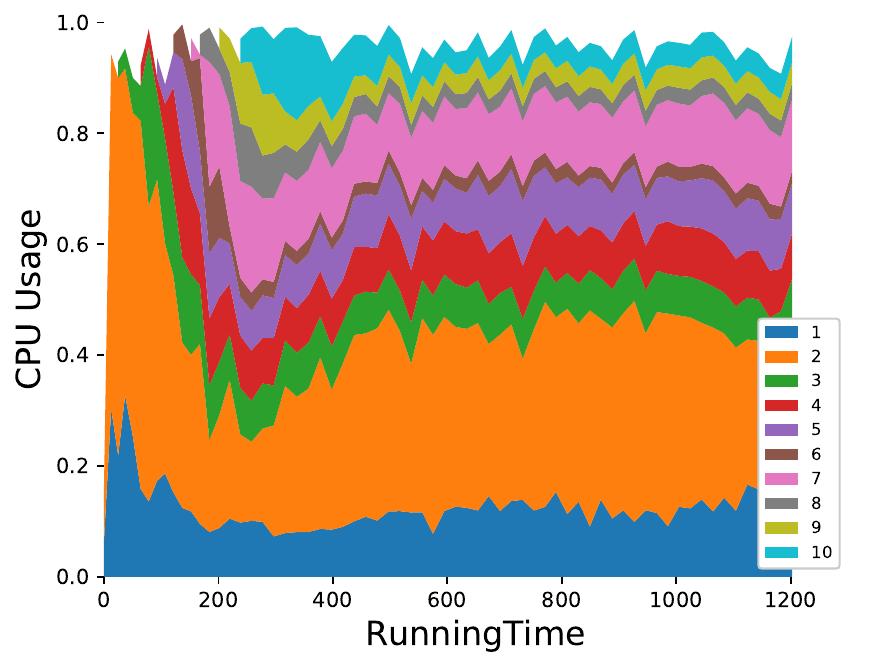}
\caption{CPU distribution: 10 jobs with varied objectives (Random schedule)}
      \label{fig:mul:random:cpu}
\end{figure}

We conduct the experiment in a single node setting, which hosts 10 running models, and the submission interval is set to [0, 300s].
A similar trend is discovered in Fig.~\ref{fig:mul:random} such that 
QoE of the system keeps getting worse from 0 to 300s. Due to a random submission schedule, 
it lacks time for \sol~ to adjust resource allocation initially.
However, given limited room, 
\sol~ dynamically configures resources with respect to their individual objectives.
For example, at time 257.07s, container-6's performance is 36.12s, and its objective is 35. Therefore,
it is classified to set $S$, which means it satisfies the predefined target.
After 300s, the overall QoE keeps approaching 0, and the number of containers in $S$ increases.
With a stable workload, \sol~ is able to react and adjust the resources to improve the QoE quickly.
Fig.~\ref{fig:mul:random:cpu} clearly indicates that the resource is not evenly distributed and, during the submission interval,
it keeps updating to adopt the dynamic workloads. In the end, it converges to a stable allocation.

\subsubsection{Cluster Environment}

\begin{figure*}[ht]
   \centering
         \begin{subfigure}[b]{0.24\textwidth}
\centering
         \includegraphics[width=\textwidth]{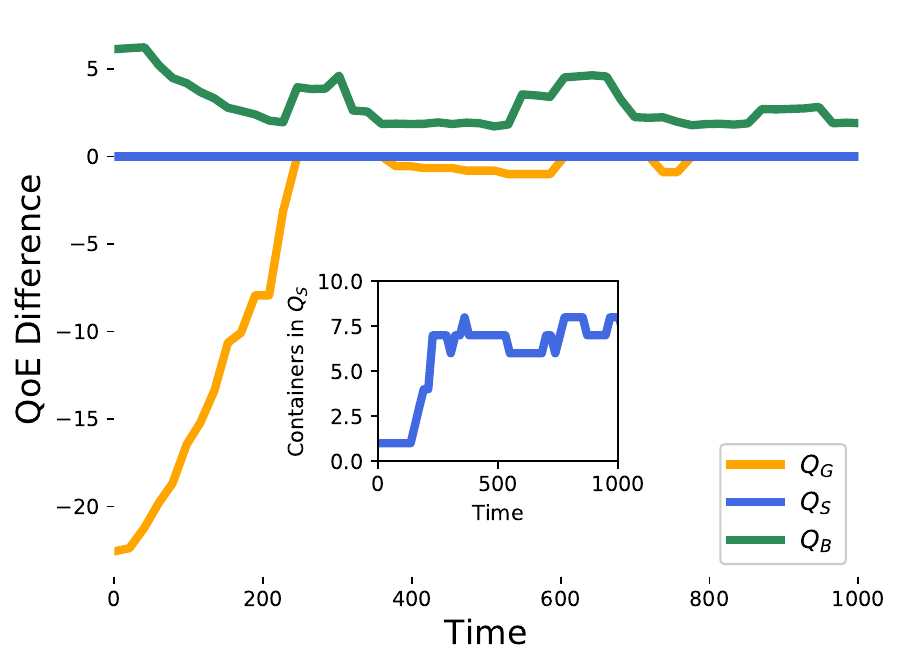}
\caption{QoE: Worker-1}
      \label{fig:cluster:worker1}
      \end{subfigure} 
      \begin{subfigure}[b]{0.24\textwidth}
\centering
         \includegraphics[width=\textwidth]{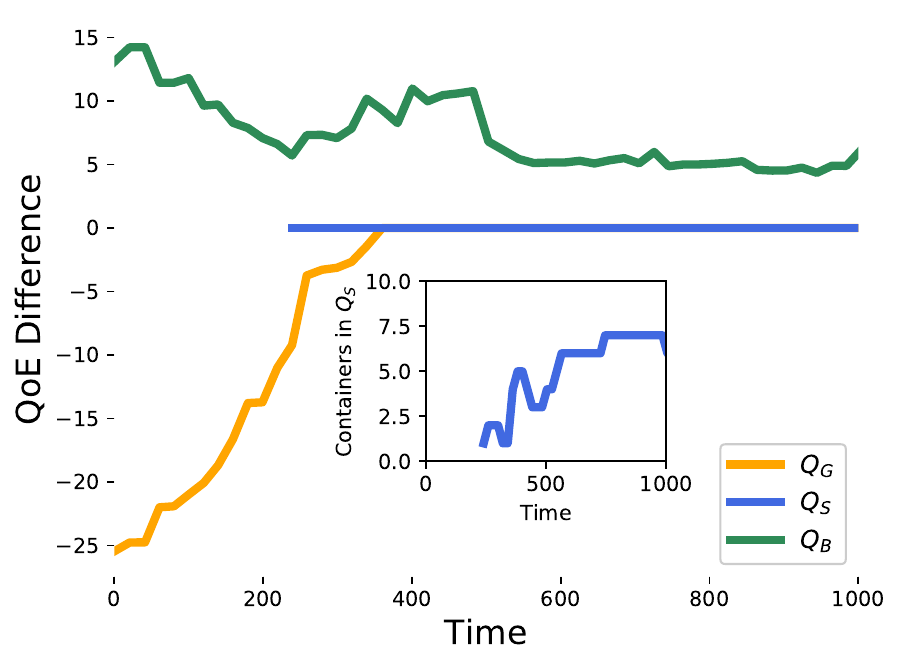}
\caption{QoE: Worker-2}
      \label{fig:cluster:worker2}
      \end{subfigure} %
      \begin{subfigure}[b]{0.24\textwidth}
\centering
         \includegraphics[width=\textwidth]{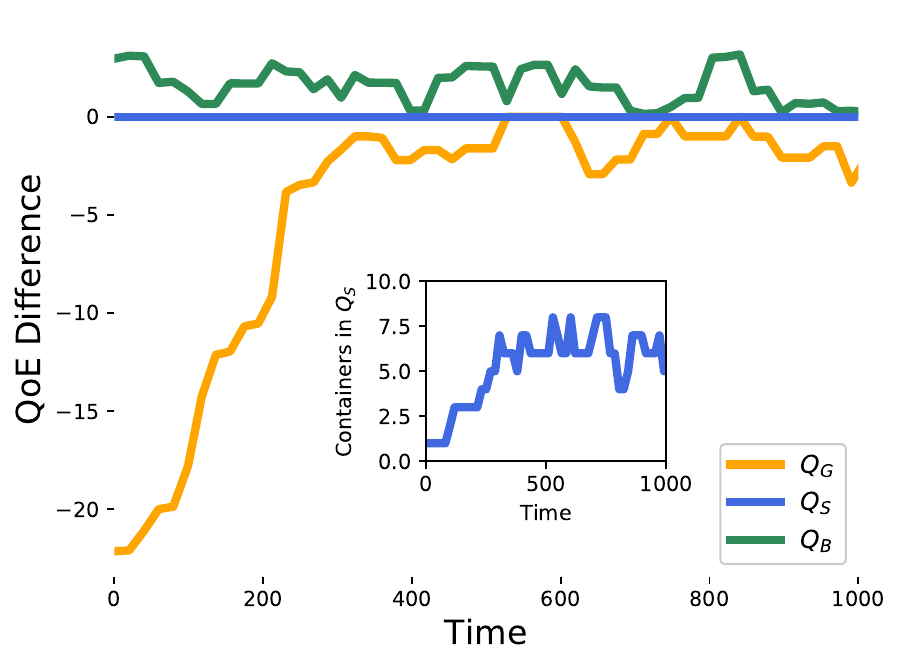}
\caption{QoE: Worker-3}
      \label{fig:cluster:worker3}
      \end{subfigure} %
      \begin{subfigure}[b]{0.24\textwidth}
	\centering
         \includegraphics[width=\textwidth]{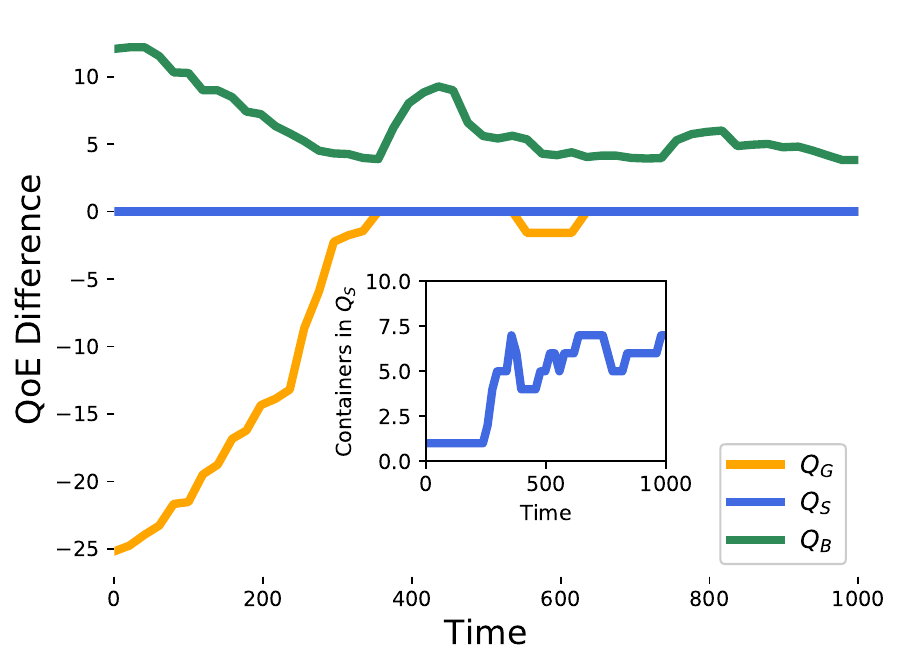}
	\caption{QoE: Worker-4}
      \label{fig:cluster:worker4}
      \end{subfigure} %
\caption{Delivered QoE in the cluster with 40 models and varied objective with \sol}    
\label{qoe:dqoes}                            
\end{figure*}

Finally, we evaluate \sol~ in a cluster environment with 4 worker nodes.
In these experiments, we launch 40 randomly selected models along with various objective values.
Fig.~\ref{qoe:dqoes} illustrates the delivered QoE of \sol.
A similar trend of results is found across all workers.
The QoE keeps improving as the algorithms execute iteratively.  
However, since the individual objectives are randomly generated, the number of containers that satisfy predefined values varies. For example, there are 8, 5, 7, 6 for Worker-1, 2, 3, 4, respectively.
In addition, when the first model appears in set $S$ is also different from each other.
For instance, on Worker-4, container-32 satisfies the objective at 20.25s, where the batch cost 65.24s and the
predefined target is 65. While on Worker-2, the first model that satisfies its objective is found at time 238.88s such that container-26 meets the requirement (68.01 v.s. 70).
When more models with unachievable objectives, e.g., Worker-3, the number of containers in $S$ is bouncing since 
\sol~ tries to make the overall QoE approach to 0, and the distribution keeps updating within underperformed models.

\begin{figure*}[ht]
   \centering
         \begin{subfigure}[b]{0.24\textwidth}
\centering
         \includegraphics[width=\textwidth]{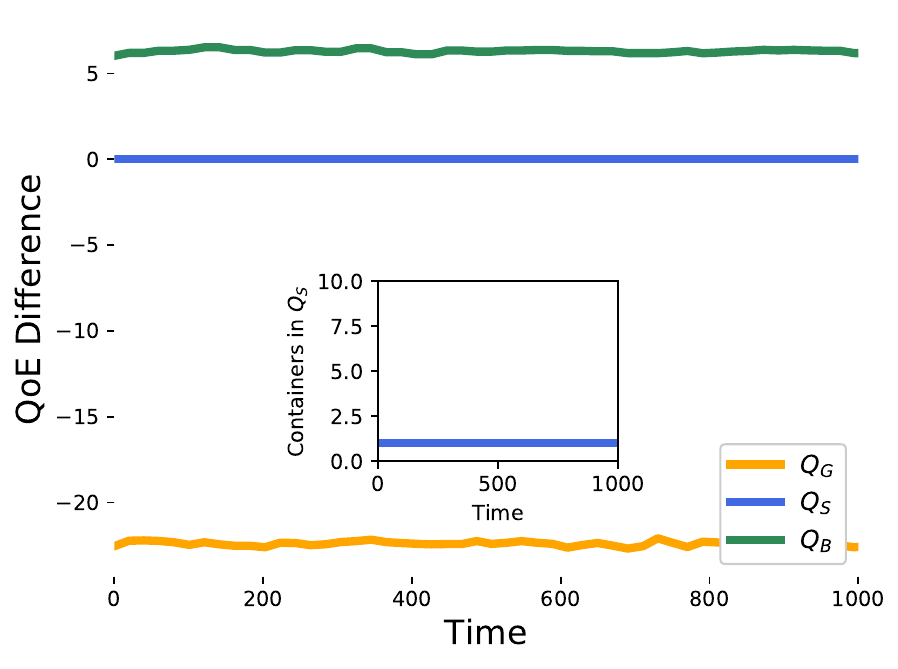}
\caption{QoE: Worker-1}
      \label{fig:ds:worker1}
      \end{subfigure} 
      \begin{subfigure}[b]{0.24\textwidth}
\centering
         \includegraphics[width=\textwidth]{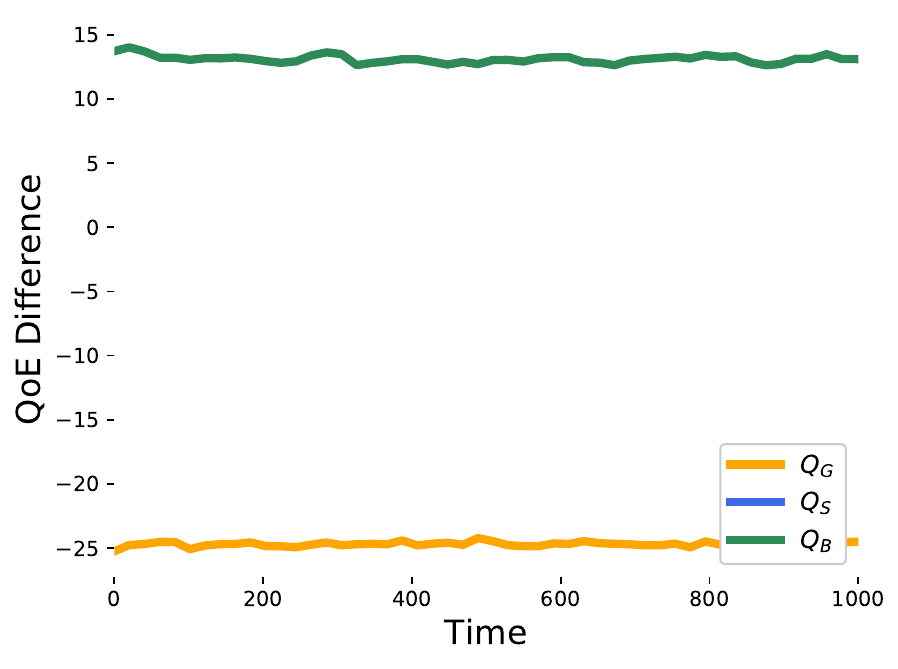}
\caption{QoE: Worker-2}
      \label{fig:ds:worker2}
      \end{subfigure} %
      \begin{subfigure}[b]{0.24\textwidth}
\centering
         \includegraphics[width=\textwidth]{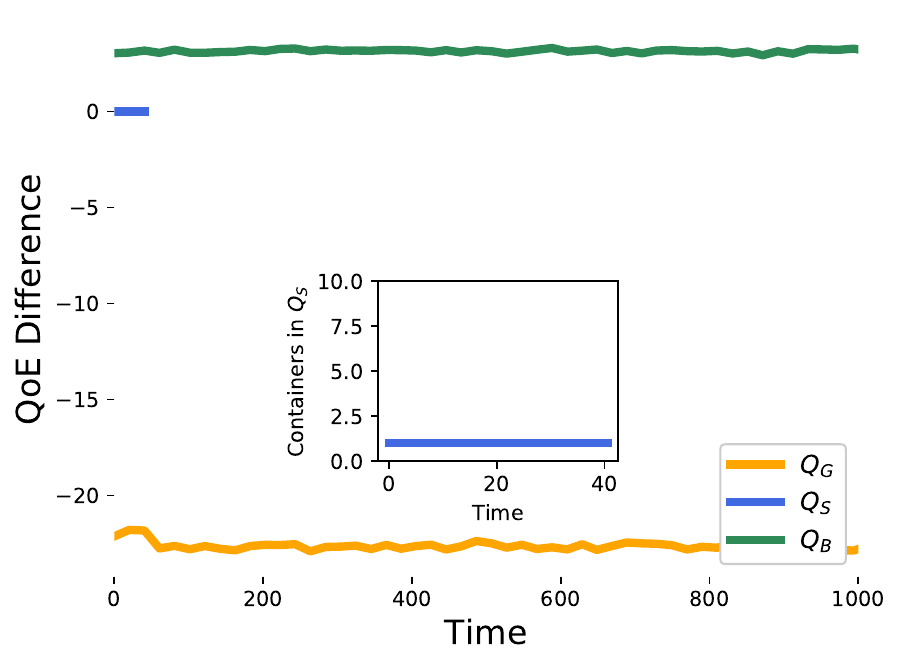}
\caption{QoE: Worker-3}
      \label{fig:ds:worker3}
      \end{subfigure} %
      \begin{subfigure}[b]{0.24\textwidth}
	\centering
         \includegraphics[width=\textwidth]{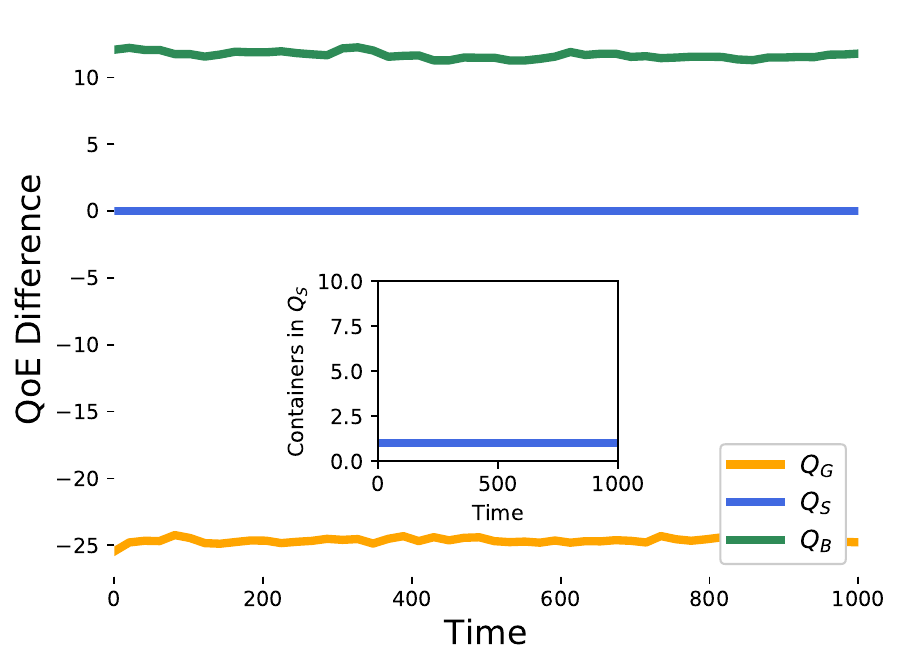}
	\caption{QoE: Worker-4}
      \label{fig:ds:worker4}
      \end{subfigure} %
\caption{Delivered QoE in the cluster with default resource management algorithms}    
\label{qoe:ds}                            
\end{figure*}

The same experiment is conducted on the original Docker Swarm platform, where default resource management algorithms are in charge of scheduling decisions. Fig.~\ref{qoe:ds} plots the results on each worker in the cluster. Clearly, due to the missing mechanism to react to the associated objective values, whether a model can satisfy the target depends on the value itself and how many concurrent running models are on the same worker.  As we can see that there is one container in set S on Worker-1, 3, and 4. On Worker-2, none of the models could meet their target. \sol~ is able to promote as many as 8x times more models to satisfy their predefined objective values,26 in \sol~ versus 3 in the default system.

\begin{figure*}[ht]
   \centering
         \begin{subfigure}[b]{0.24\textwidth}
\centering
         \includegraphics[width=\textwidth]{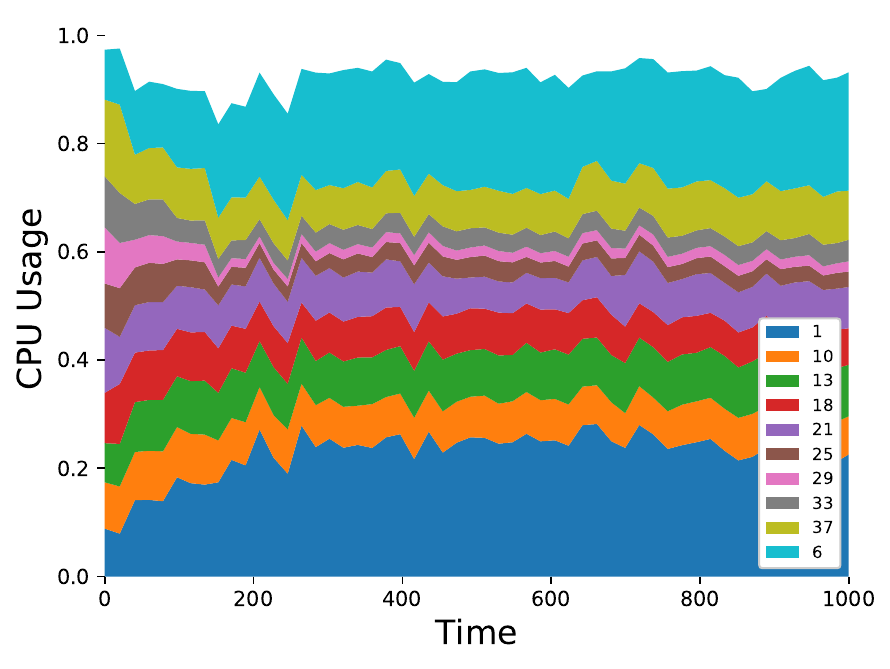}
\caption{QoE: Worker-1}
      \label{fig:cluster:worker1:cpu}
      \end{subfigure} 
      \begin{subfigure}[b]{0.24\textwidth}
\centering
         \includegraphics[width=\textwidth]{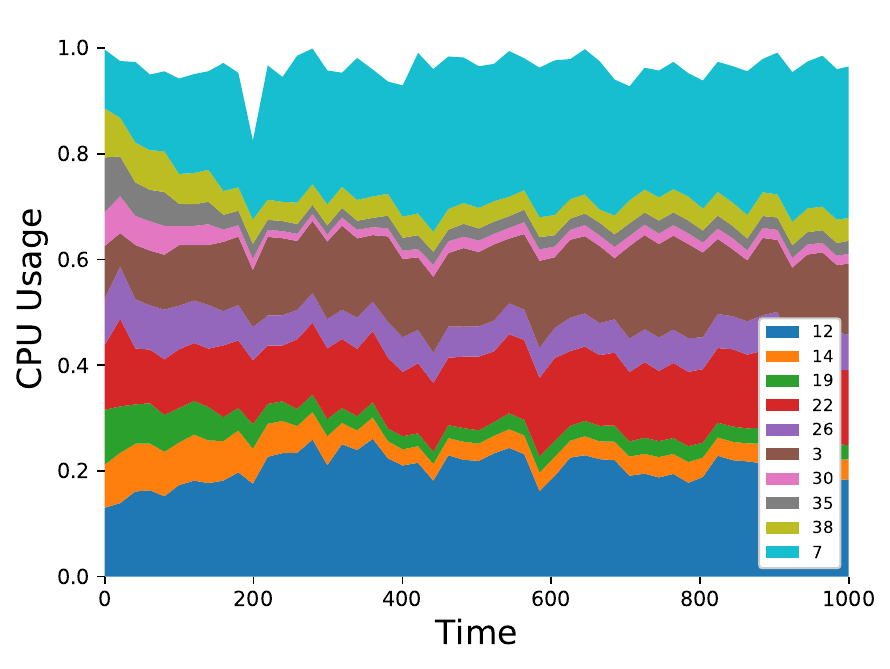}
\caption{QoE: Worker-2}
      \label{fig:cluster:worker2:cpu}
      \end{subfigure} %
      \begin{subfigure}[b]{0.24\textwidth}
\centering
         \includegraphics[width=\textwidth]{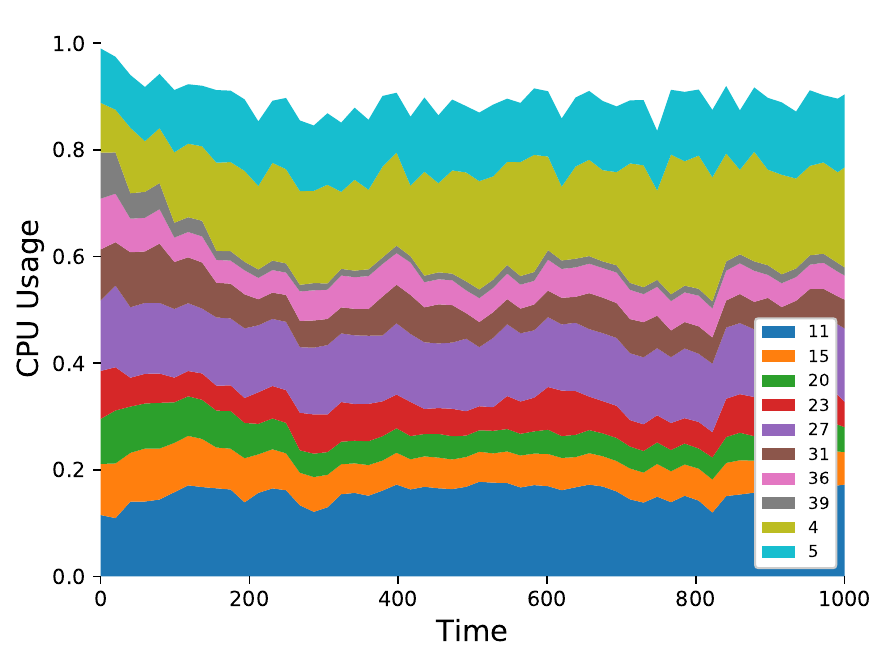}
\caption{QoE: Worker-3}
      \label{fig:cluster:worker3:cpu}
      \end{subfigure} %
      \begin{subfigure}[b]{0.24\textwidth}
	\centering
         \includegraphics[width=\textwidth]{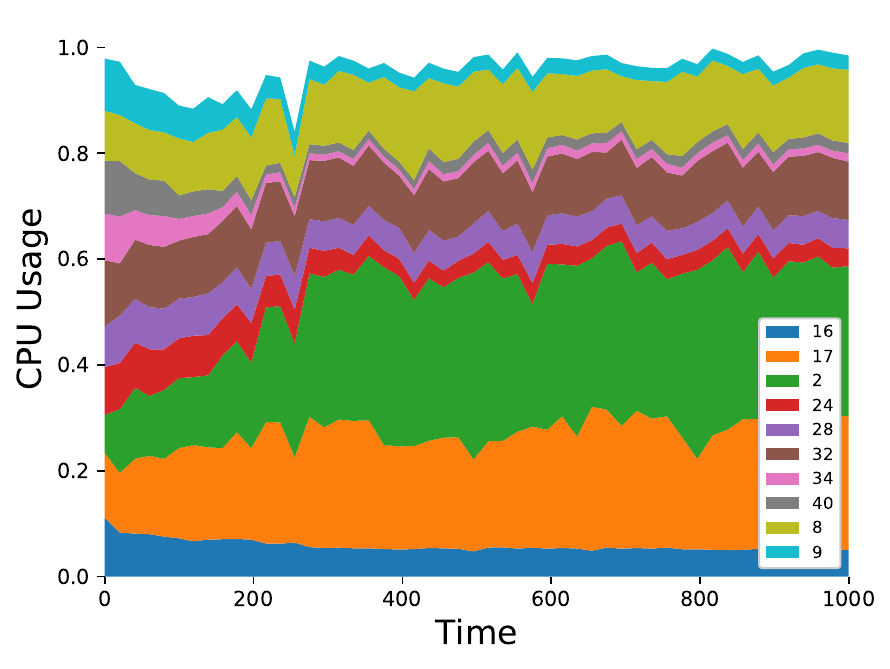}
	\caption{QoE: Worker-4}
      \label{fig:cluster:worker4:cpu}
      \end{subfigure} %
\caption{CPU distribution in the cluster with 40 models and varied objective with \sol}    
\label{cpu:dqoes}                            
\end{figure*}

\begin{figure*}[ht]
   \centering
         \begin{subfigure}[b]{0.24\textwidth}
\centering
         \includegraphics[width=\textwidth]{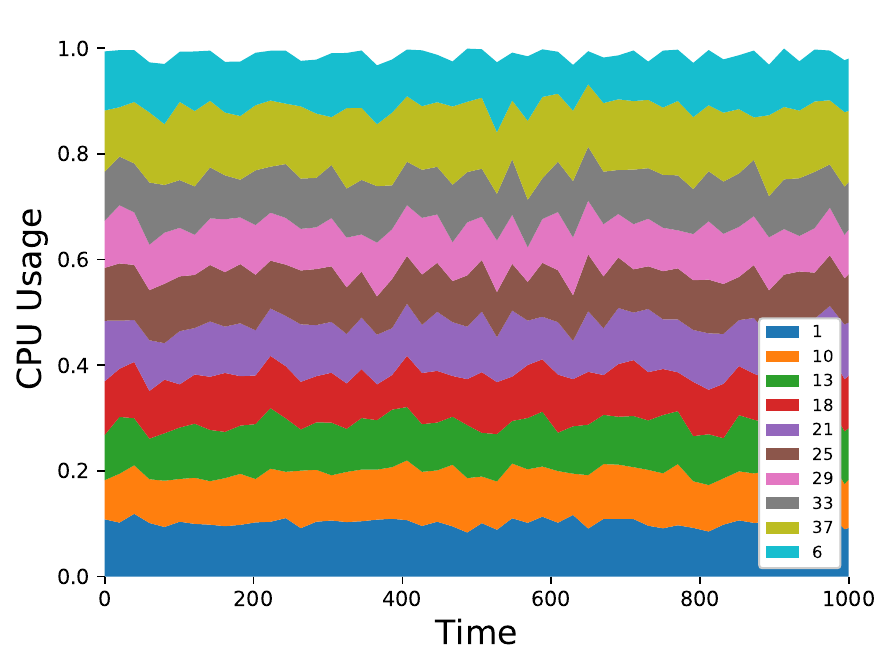}
\caption{QoE: Worker-1}
      \label{fig:ds:worker1:cpu}
      \end{subfigure} 
      \begin{subfigure}[b]{0.24\textwidth}
\centering
         \includegraphics[width=\textwidth]{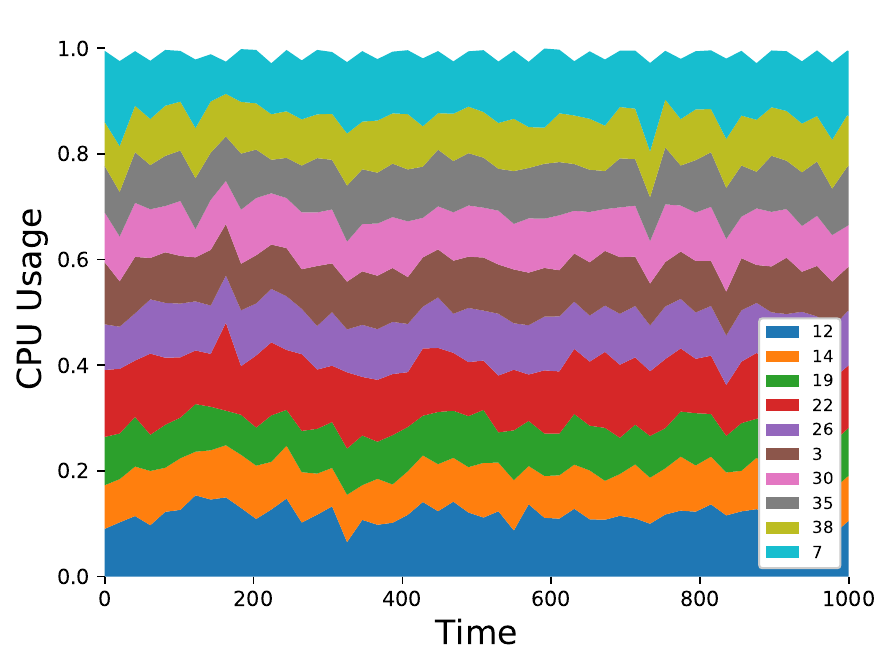}
\caption{QoE: Worker-2}
      \label{fig:ds:worker2:cpu}
      \end{subfigure} %
      \begin{subfigure}[b]{0.24\textwidth}
\centering
         \includegraphics[width=\textwidth]{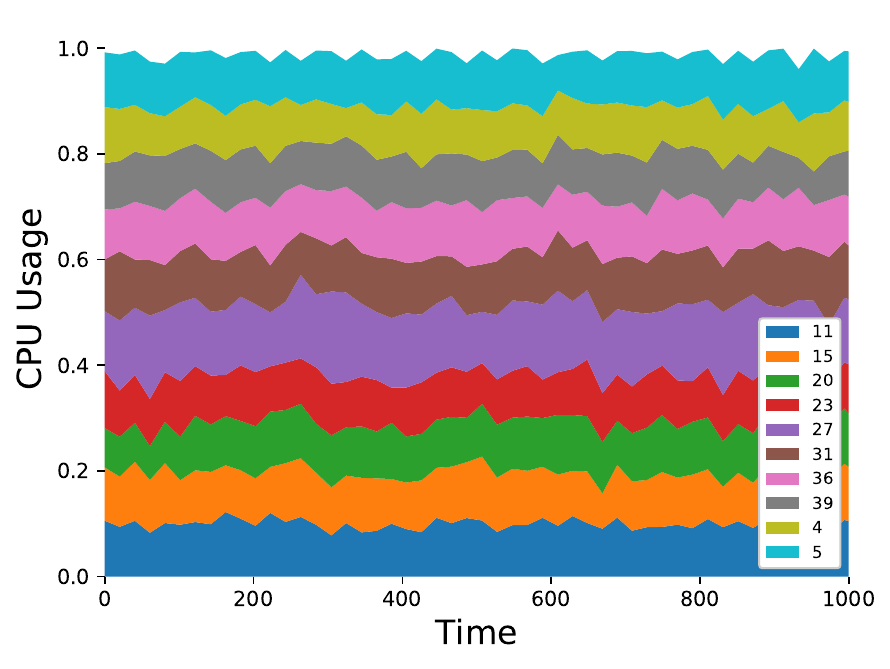}
\caption{QoE: Worker-3}
      \label{fig:ds:worker3:cpu}
      \end{subfigure} %
      \begin{subfigure}[b]{0.24\textwidth}
	\centering
         \includegraphics[width=\textwidth]{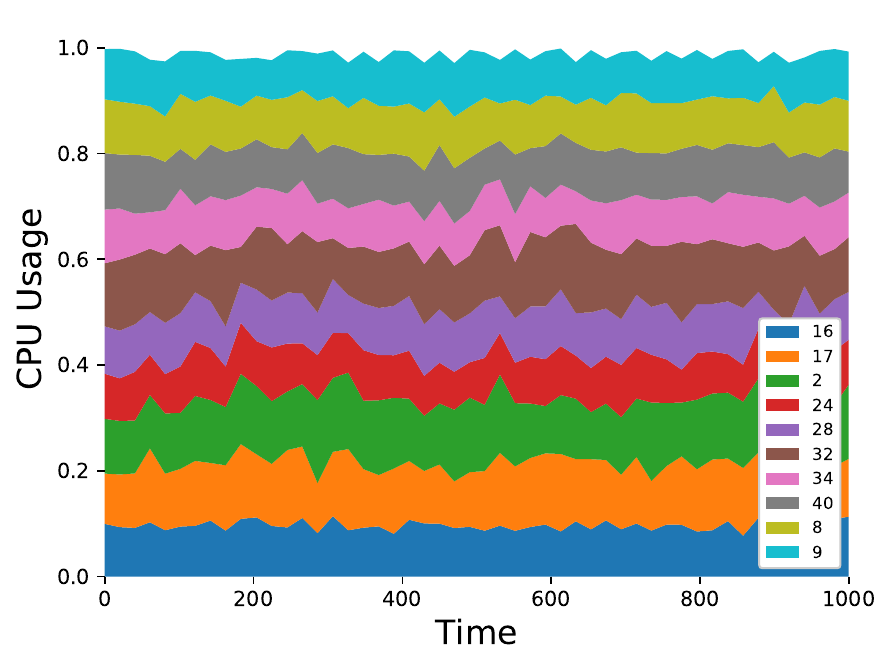}
	\caption{QoE: Worker-4}
      \label{fig:ds:worker4:cpu}
      \end{subfigure} %
\caption{CPU distribution in the cluster with default resource management algorithms}    
\label{cpu:ds}                            
\end{figure*}

Fig.~\ref{cpu:dqoes} and Fig.~\ref{cpu:ds} present the CPU distribution of the two experiments.
Comparing them, the improvement is achieved by dynamically adjusting resource allocation at runtime with respect for each objective value. When there is an unachievable one, \sol~ attempts to utilize the released resources fully from the models with an achievable objective. 
With a default system, however, the resource can only be evenly distributed, where each individual container gets its equal share of the resource.

\section{Conclusion}
\label{con}

In this work, we present \sol~ a differentiate quality of experience based scheduler
for containerized deep learning applications.
It enables user-specified objectives to the cloud system and attempts to meet multiple clients' specifications 
in the same cluster.
When deploying a learning model in the cloud, \sol~ accepts a targeted QoE value from each client. 
This value is analyzed by \sol, which would increase or decrease its resource limit in order to
achieve the target.
When multiple models reside in a cluster, \sol~ is able to respect to their individual QoE objectives and 
approach user-specified QoEs through dynamical adjustment on the resource allocation at runtime.
\sol~ is implemented as a plugin to Docker Swarm. Based on extensive, cloud executed experiments, it demonstrates
its capability to react to different workload patterns and achieves up to 8x times 
more satisfied models as compared to the existing system.

Given the limited on-board resource on the worker, to further improve \sol~ in a cluster environment, horizontal scaling is necessary. We will investigate container placement and migration strategies with respect to run-time states of worker nodes. By default, Docker Swarm places containers based on the number of containers on each worker and assumes each container allocating an equal share of resources. However, when there is an underperformed job on a worker that is in a stable state, the system should avoid putting more workload on this worker, which may result in an increased number of underperformed jobs. We plan to enhance \sol~ with novel container placement and migration algorithms to balance workload according to clients' specifications. Additionally, the fairness between containers and workers as well as other resource types, e.g., memory and network, should be considered.

\bibliographystyle{IEEEtran}
\bibliography{sections/routing}


\vskip -2\baselineskip plus -1fil

\begin{IEEEbiography}[{\includegraphics[width=0.9in,height=1in,clip,keepaspectratio]{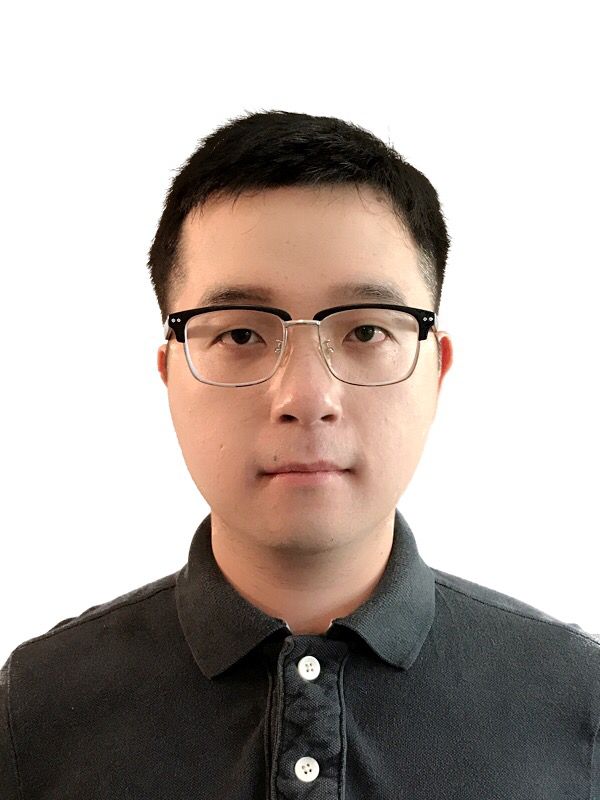}}]{Ying Mao} is an Assistant Professor in the Department of Computer and Information Science at Fordham University in the New York City. He received his Ph.D. in Computer Science from the University of Massachusetts Boston. He was a Fordham-IBM research fellow. His research interests mainly focus on the fields of advanced computing systems, virtualization, resource management, data-intensive platforms and quantum-based applications. He is a member of the IEEE.
\end{IEEEbiography}

\vskip -2\baselineskip plus -1fil

\begin{IEEEbiography}[{\includegraphics[width=0.9in,height=1in,clip,keepaspectratio]{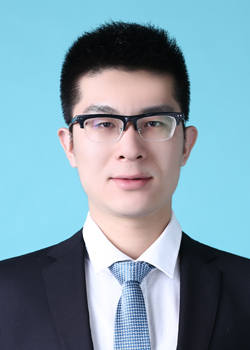}}]{Weifeng Yan} was a graduate student at Fordham University. He earned his Master of Science in Data Science degree in 2020. Additionally, he holds a Bachelor’s degree in Mathematics at Shanghai Jiao Tong University. 
His research  interests focus on machine learning, big data, and data mining. He is proficient in coding with deep learning frameworks and data analytic platforms.
\end{IEEEbiography}

\vskip -2\baselineskip plus -1fil

\begin{IEEEbiography}[{\includegraphics[width=0.9in,height=1in,clip,keepaspectratio]{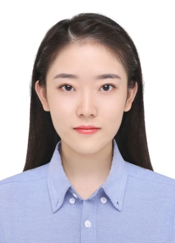}}]{Yun Song} was a graduate student at Fordham University, where  
she received the M.S. degree in Data Science in 2020. 
Previously, she received the B.S. degree in Management from Sichuan University in 2018. Her research interests are in deep learning, cloud computing and resource management.
\end{IEEEbiography}

\vskip -2\baselineskip plus -1fil

\begin{IEEEbiography}[{\includegraphics[width=0.9in,height=1in,clip,keepaspectratio]{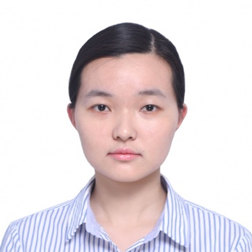}}]{Yue Zeng} was a research assistant at Fordham University in New York City. She received her Master of Science in Data Science degree in May, 2020. Her research interests mainly about machine learning and data mining analysis.
\end{IEEEbiography}

\vskip -2\baselineskip plus -1fil

\begin{IEEEbiography}[{\includegraphics[width=0.9in,height=1in,clip,keepaspectratio]{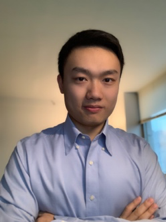}}]{Ming Chen} was a graduate student in the department of computer and information science at Fordham University in New York City. He earned his Master of Science in Data Science degree in May 2020. His research interests are data mining, cloud computing, distributed system.
\end{IEEEbiography}

\vskip -2\baselineskip plus -1fil

\begin{IEEEbiography}[{\includegraphics[width=0.9in,height=1in,clip,keepaspectratio]{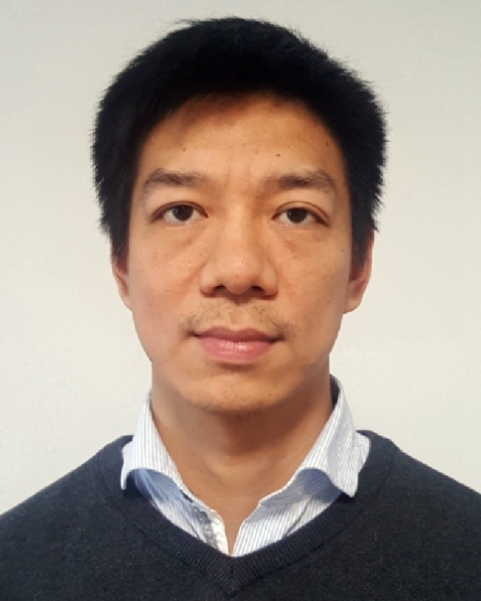}}]{Long Cheng}  is a Full Professor in the School
of Control and Computer Engineering at North
China Electric Power University in Beijing, and
also a Visiting Professor at Insight SFI Research
Centre for Data Analytics in Dublin. He received
the B.E. from Harbin Institute of Technology,
China in 2007, M.Sc from University of Duisburg-Essen, Germany in 2010 and Ph.D from National
University of Ireland Maynooth in 2014. He was
an Assistant Professor at Dublin City University,
and a Marie Curie Fellow at University College
Dublin. He also has worked at organizations such as Huawei Technologies Germany, IBM Research Dublin, TU Dresden and TU Eindhoven.
He has published more than 60 papers in journals and conferences like
TPDS, TON, TC, TSC, TASE, TCAD, T-ITS, TCC, TBD, TVLSI, JPDC,
IEEE Network, CIKM, ICPP, CCGrid and Euro-Par etc. His research
focuses on distributed systems, deep learning, cloud computing and
process mining. Prof Cheng is a Senior Member of the IEEE and an
associate editor of the Journal of Cloud Computing.
\end{IEEEbiography}

\vskip -2\baselineskip plus -1fil

\begin{IEEEbiography}[{\includegraphics[width=0.9in,height=1in,clip,keepaspectratio]{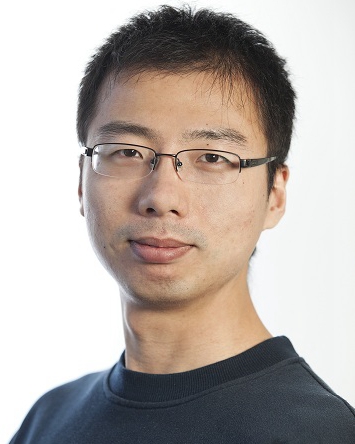}}] {Qinzhi Liu} is a Lecturer at the Information Technology Group, Wageningen University, The Netherlands. He received a B.E. and a M.Eng. from Xidian University, China in 2005 and 2008 respectively. He received a M.Sc. (with cum laude) and a Ph.D. from Delft University of Technology, The Netherlands in 2011 and 2016 respectively. He was a Postdoctoral Researcher at the System Architecture and Networking Group, Eindhoven University of Technology, The Netherlands from 2016 to 2019. His research interests include Internet of Things and machine learning.
\end{IEEEbiography}



\end{document}